\newcommand{\sqrtsnn}{\sqrt{s_{_{\rm NN}}}}

\documentclass{PoS}

\title{Direct Photon and Heavy Quark Jet Production at the LHC}

\ShortTitle{Direct Photon and Heavy Quark Jet Production at the LHC}

\author{T. Stavreva\\
        Laboratoire de Physique Subatomique et de Cosmologie, UJF, CNRS/IN2P3,
\\
INPG, 53 avenue des Martyrs, 38026 Grenoble, France\\
        E-mail: \email{stavreva@lpsc.in2p3.fr}}

\abstract{The associated production of direct photons and heavy quarks at the LHC is presented.  Predictions for the nuclear modification factor to the cross-section at ALICE are shown.  It is demonstrated that this process is a great probe of the gluon and heavy quark nuclear PDFs, as over 80$\%$ of its nPDF dependence at NLO comes from those nPDFs.  Therefore measurements of this process will provide an excellent constraint on the gluon nPDF, and will distinguish between different nPDF sets currently out on the market. }

\FullConference{35th International Conference of High Energy Physics - ICHEP2010,\\
		July 22-28, 2010\\
		Paris France}

\begin{document}

Direct photons can serve as an excellent probe of the structure of the proton due to their pointlike electromagnetic coupling to quarks and due to the fact that they escape confinement.  However the information that can be obtained about the underlying subprocesses and separate parton distribution functions (PDFs) by looking at direct photons alone can be somewhat limited. This can be rectified by  investigating the associated production of  direct  photons  with heavy quarks (charm or bottom), thereby, providing valuable information on the gluon and heavy quark PDFs (for more details 
see Ref. \cite{Stavreva:2009vi}). 
This study can also be extended to high-energy proton-nucleus collisions (p-A), where one can use $\gamma + Q$ production to investigate the structure of the nucleus as well and constrain the large error associated with the gluon nuclear PDF (nPDF).  
The precise knowledge of the gluon nPDF ($g^{p/A}(x,Q)$) is necessary to avoid the otherwise significant uncertainty associated with it in the theoretical predictions for hard processes in $A-A$ collisions.
%%%%%%
\begin{figure}[t] 
\begin{center}
\includegraphics[angle=-90,scale=0.25,trim=3.2cm 0cm 1cm 0cm,clip]{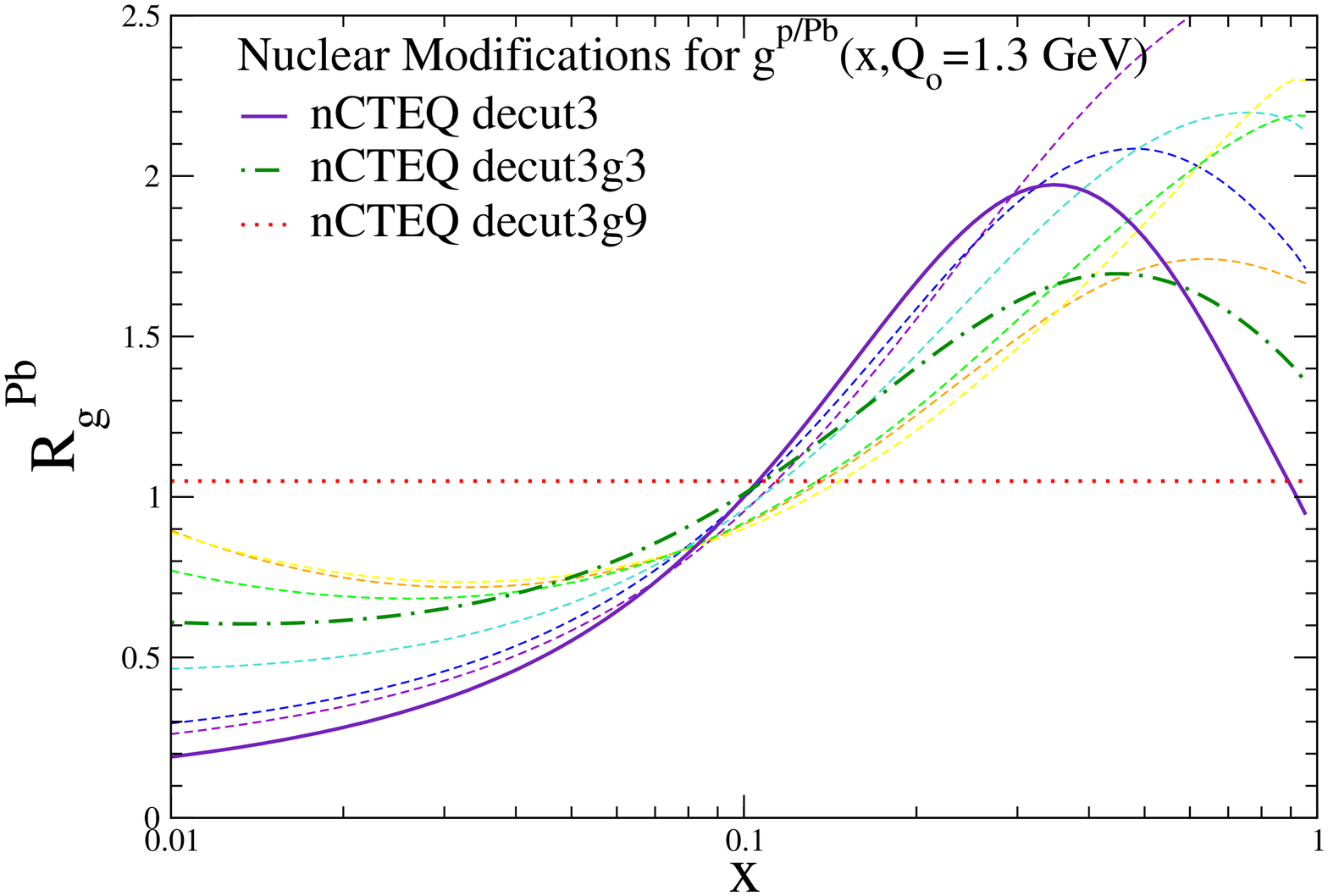} 
\includegraphics[angle=-90,scale=0.25,trim=3.2cm 0cm 1cm 0cm,clip]{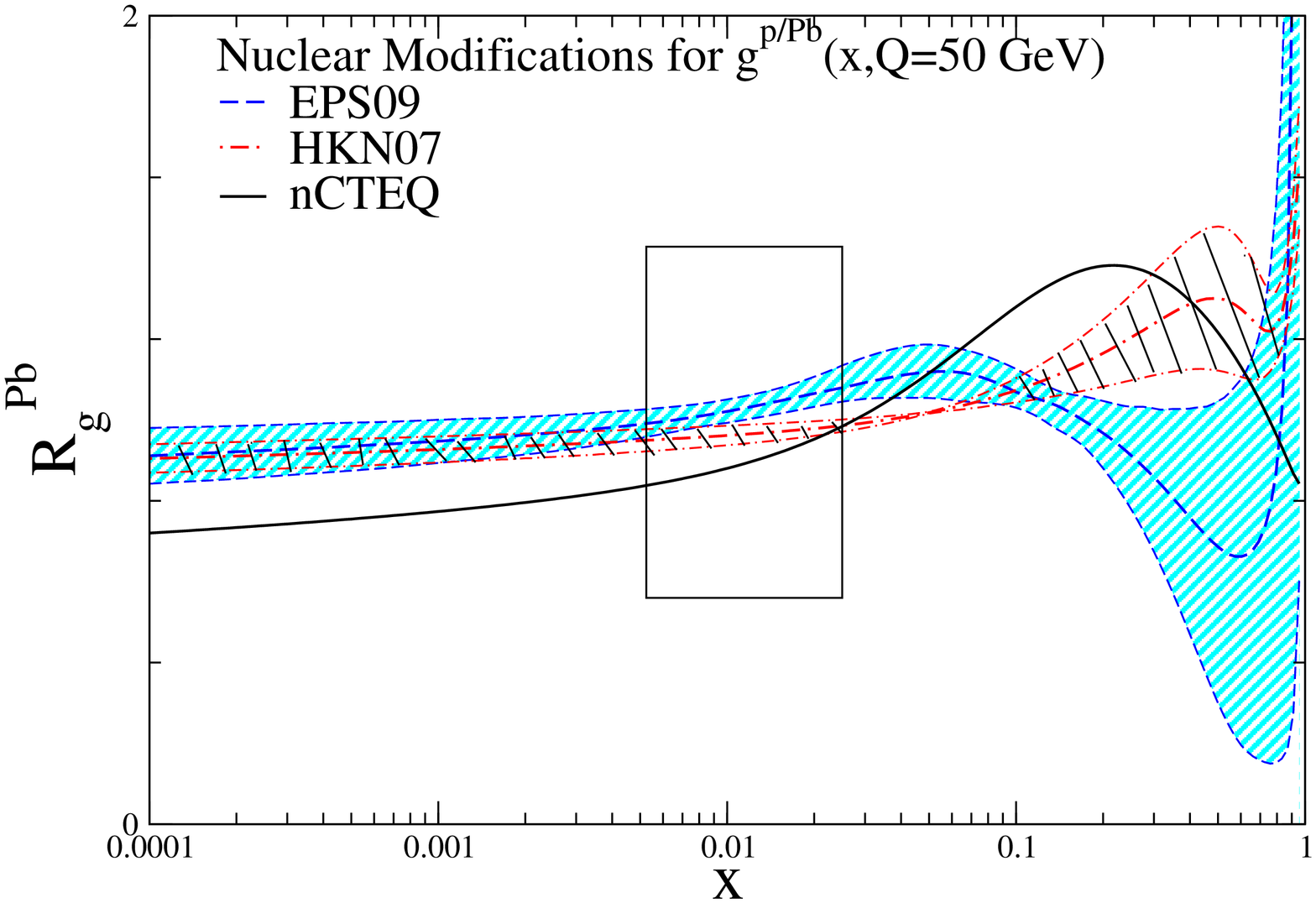}
\caption{Left: nPDF ratio $R_g^{Pb}$ at a scale $Q_0=1.3$~GeV showing the spread of different equally acceptable nCTEQ sets -- fits from top to bottom: decut3g9, decut3g5, decut3g7, decut3g8, decut3g3, decut3g4, decut3g2, decut3g1, decut3g. Right: Nuclear modifications ${R_g^{Pb}}=g^{p/Pb}(x,Q)/g^{p}(x,Q)$ for lead at $Q=50$ GeV using nCTEQ decut3 (solid, black line),
EPS09 (dashed, blue line) + error band, HKN07 (dash-dotted, red line) + error band.
The box exemplifies the $x$-region probed at the LHC ($\sqrtsnn=8.8$~TeV).}
\label{fig:gluon}
\end{center}
\end{figure}

The nuclear gluon distribution is only very weakly constrained by NMC data on $F_2^D(x,Q^2)$ and $F_2^{Sn}/F_2^C(x,Q^2)$ 
\footnote{This is true for most nPDF fits, while EPS09 also includes 
data on $\pi^0$ production at RHIC.}.
The large uncertainty in the nuclear gluon as well as the variations in  
predictions of different fits (nCTEQ \cite{nCTEQ1,nCTEQ2,Kovarik:2010uv}, HKN07 \cite{Hirai:2007sx}, 
EPS09 \cite{EPS}) is presented in Fig.\ \ref{fig:gluon} in the form of the gluon nuclear 
modification factor $R_g(x,Q)=g^{p/Pb}(x,Q)/g^p(x,Q)$. 
It is quite clear that there is a strong need for data constraining the gluon nPDF. 
In the general framework (disregarding the possibility of intrinsic charm presence), the charm (and respectively bottom) distribution is solely based on the gluon PDF via the DGLAP evolution equations. 
Therefore a process such as $\gamma + Q$ production, sensitive to those distributions will be excellent for constraining the gluon nPDF. 

\begin{figure} 
\begin{center}
   \includegraphics[angle=-90,scale=0.26,trim=3.2cm 0cm 1cm 0cm,clip]{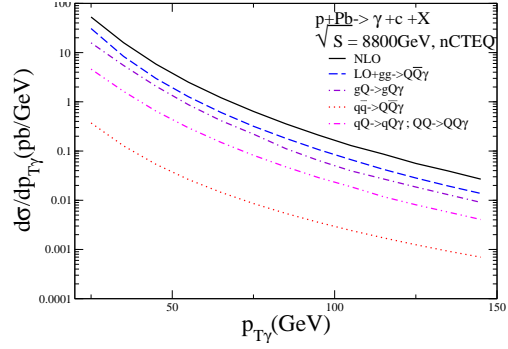}
\caption{Subprocess contributions to $d \sigma^{\gamma + c}/dp_{T\gamma}$, NLO (solid black line), LO$+ gg \rightarrow Q\bar Q\gamma$ (dashed blue line), $gQ\rightarrow gQ\gamma$ (dash-dotted purple line), $q\bar q \rightarrow Q\bar Q \gamma$ (dotted red line), 
$qQ \rightarrow qQ \gamma$; $QQ \rightarrow QQ \gamma$ (dash-dot-dotted magenta line).}
\label{fig:EMCalparts}
\end{center}
\end{figure}

\begin{figure} 
\begin{center}
    \includegraphics[angle=-90,scale=0.25,trim=3.2cm 0cm 1cm 0cm,clip]{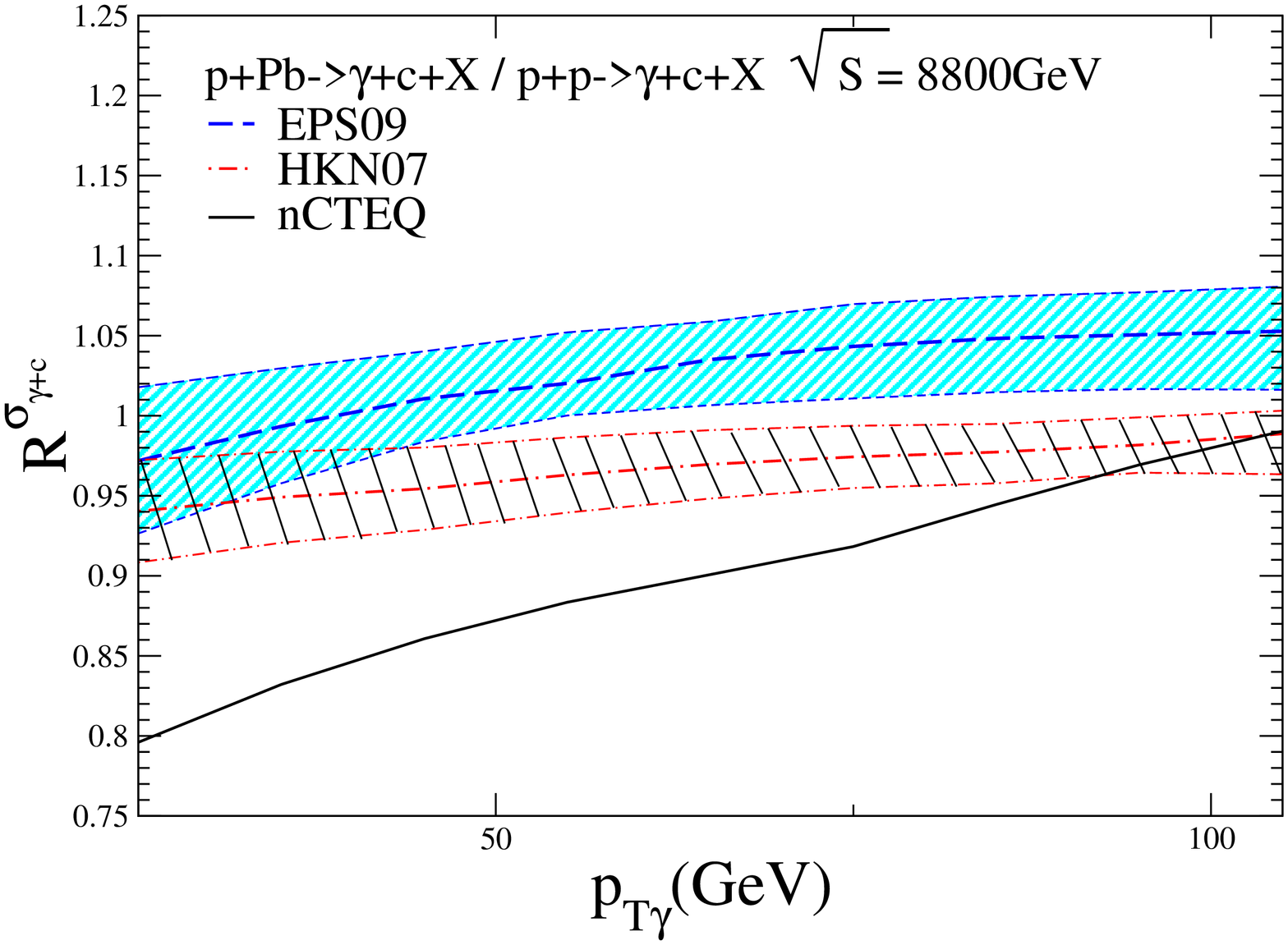}
    \includegraphics[angle=-90,scale=0.25,trim=3.2cm 0cm 1cm 0cm,clip]{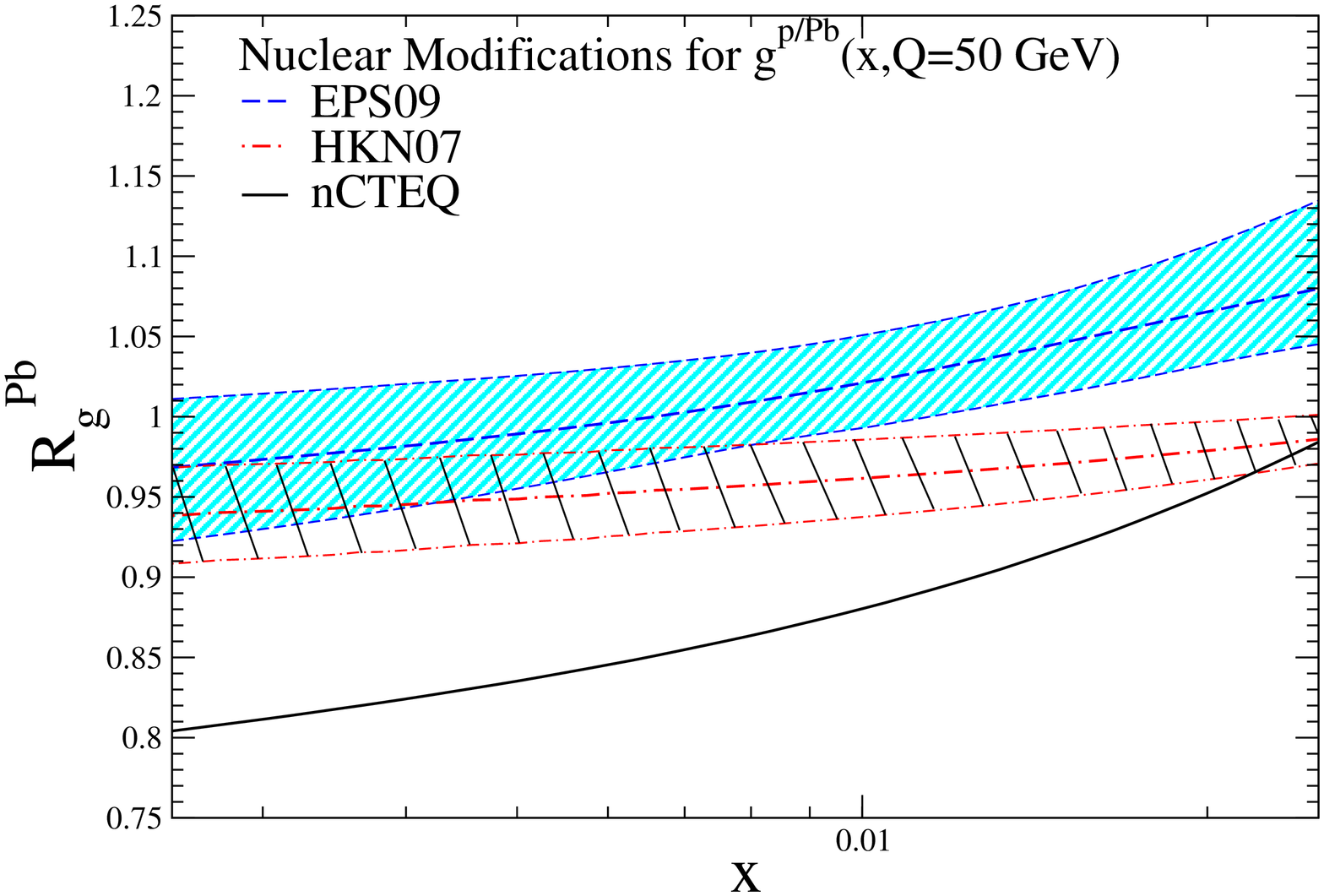}
\caption{Left: nuclear production ratio of $\gamma+c$ cross-section at LHC within ALICE PHOS acceptances, 
using nCTEQ (solid black line), EPS09 (dashed blue line) + error band, HKN07 (dash-dotted red line) + error band. 
Right: $R_g^{Pb}$ ratio as a function of $x$, in the $x$ region probed at the LHC.}
\label{fig:nRatio}
\end{center}
\end{figure}

In Fig.\ \ref{fig:EMCalparts} the subprocess contributions to the $\gamma + c$ differential cross-section are presented. There the dominating subprocesses at $\sqrtsnn=8.8$~TeV  for $p-A$ collisions are $gg$ and $gQ$ initiated, 
therefore $\gamma + c$ is a very useful process for constraining the gluon nPDF 
{\footnote {The same PDF dependence is true for $\gamma + b$ production as well.}}.  In Fig.\ \ref{fig:nRatio} (left) the nuclear modification factor, 
$R^{\sigma^{\gamma+c}}={{d\sigma/dp_{T\gamma}(pA)} \over {A_{Pb}d\sigma/dp_{T\gamma}(pp)}}$, to the direct photon 
and charm cross-section is shown. In Fig.\ \ref{fig:nRatio} (right) we show the nuclear modifications of the gluon distribution in a lead nucleus  
for the typical $x$-region probed at the LHC.   
It can be clearly seen by comparing the left and right side of Fig.\ \ref{fig:nRatio} that the nuclear 
modification factor to the cross-section, $R^{\sigma^{\gamma+c}}$, follows closely the nuclear modifications 
for the gluon nPDF ($R_g$), in the region of $x$ probed at the LHC; for more details see \cite{Stavreva:2010mw}.   
Therefore  an appropriate measurement of this process will be able to distinguish 
between the three different nPDF sets.  
We can conclude that this process is an excellent candidate for constraining the gluon nuclear distribution. 


\end{document}